\documentclass[10pt]{article}
\begin{document}
%%%%%%%%%%%%%%%%%%%%
\title{{\bf Hawking radiation in GHS and non-extremal 
D1-D5 blackhole via covariant 
anomalies}}
%%%%%%%%%%%%%%%%%%%%
\author{
{Sunandan Gangopadhyay$^{}$\thanks{sunandan@bose.res.in}} and 
{Shailesh Kulkarni$^{ }$\thanks{shailesh@bose.res.in} }\\
S.~N.~Bose National Centre for Basic Sciences,\\JD Block, 
Sector III, Salt Lake, Kolkata-700098, India\\[0.3cm]
}
\date{}

\maketitle

%%%%%%%%%%%%%%%%
\begin{abstract}
%%%%%%%%%%%%%%%%
\noindent We apply the method of Banerjee and Kulkarni 
(arXiv:0707.2449, [hep-th]) to provide
a derivation of Hawking radiation from the GHS (stringy) blackhole
which falls in the class of the most general
spherically symmetric blackholes ($\sqrt{-g}\neq 1$) and also
the non-extremal $D1-D5$ blackhole using 
only covariant gravitational anomalies. 
\\[0.3cm]
\\[0.3cm]
%{\bf PACS:} 11.10.Nx 

\end{abstract}
%%%%%%%%%%%%%%%%%%%%%%%%%%%%%%%%%%%%%%%%%%%%%%%%%%%%%%%%%%%%%%%

\noindent {\it{Introduction :}}

\noindent Hawking radiation is one of the most prominent quantum effect 
that arises for quantum fields in a background spacetime with an event
horizon. The radiation is found to have a spectrum 
with Planck distribution giving
the blackholes one of its thermodynamic properties.
There are several derivations and all of them take the quantum
effect of fields in blackhole backgrounds into account in various ways. 
The original derivation by Hawking \cite{Hawking:sw} \cite{Hawking:rv} 
calculates the Bogoliubov coefficients
between the in and out states of fields in a blackhole background. 
A tunneling picture \cite{Parikh:1999mf, tunneling}
is based on pair creations of particles and antiparticles near the horizon
and calculates WKB amplitudes for classically forbidden paths. 
A common property in these derivations is the universality of the radiation: 
i.e. Hawking radiation is determined universally by the horizon properties
(if we neglect the grey body factor induced by the effect of scattering
outside the horizon). 

\noindent Another approach to the Hawking radiation is to calculate the
energy-momentum (EM) tensor in the blackhole backgrounds.
Classically, the EM tensor of any field is expected to be covariantly
conserved in a curved background. However, quantum mechanically this is
not always the case. For example, for a chiral scalar field in 
$(1+1)$-dimensional curved spacetime the covariant derivative of the
EM tensor reads
\begin{eqnarray}
\nabla_{\mu}{T^{\mu}}_{\nu}&=&\frac{1}{96\pi\sqrt{-g}}
\epsilon^{\beta\delta}\partial_{\delta}\partial_{\alpha}
{\Gamma^{\alpha}}_{\nu\beta}
\label{anomaly}
\end{eqnarray}
the right hand side being the consistent gravitational 
anomaly in that spacetime (\cite{bert, bert1, witt, fuji}).
\noindent Under certain simplifying assumptions, it was shown by
Christensen and Fulling \cite{full}, that the above anomaly can be
interpreted as a flux of radiation, which quantitatively agrees with
the Hawking flux \cite{Hawking:sw, Hawking:rv}, 
from a horizon in that spacetime. 

\noindent Recently, the above idea was resurrected by Robinson
and Wilczek who showed (without many of the previous
assumptions) that the above result was valid for a variety of
spacetimes (\cite{rw}). The method was soon extended to the case of 
charged blackholes \cite{iso}. Further applications of this
approach may be found in \cite{Muratasoda1}-\cite{das1}. 
The basic idea in \cite{rw,iso} is
that the effective theory near the horizon becomes 
two-dimensional and chiral. This chiral theory is anomalous. 
Using the form for two
dimensional consistent gauge/gravitational anomaly Hawking fluxes
are obtained. However the boundary condition necessary to fix the
parameters are obtained from a vanishing of covariant current and 
energy-momentum tensor at the horizon. A more conceptually cleaner
and economical approach based on cancellation 
of covariant (gauge/gravitational) anomaly
has been discussed in \cite{rb}. Since the boundary condition involved
the vanishing of covariant current/energy-momentum tensor at the horizon,
it is more natural to make use of covariant expressions for
gauge and gravitational anomaly.
The generalization of this approach to higher
spin field has been done in \cite{isorecent}    
The spacetimes considered in this method 
included many of the known spherically symmetric spacetimes.
Also, we would like to point
 out that an alternative derivation of Hawking flux based  on effective action
using only covariant anomaly has been discussed in \cite{rb2}.

\noindent In this paper, we adopt the method in (\cite{rb}) to discuss
 Hawking radiation from blackhole backgrounds in string theory.
First we discuss the Garfinkle-Horowitz-Strominger (GHS) 
blackhole which is an example of the 
most general spherically symmetric blackhole spacetime ($\sqrt{-g}\neq1$)
\cite{gibb, ghs} and then we study the non-extremal D1-D5 blackhole
% The discussion is specifically done for Hawking radiation from Garfinkle-Horowitz-
%Strominger (GHS) blackhole in string theory which is an example
%of the most general spherically symmetric blackhole spacetime 
\cite{D1}.\\

\noindent {\it{Hawking radiation from GHS blackhole :}}\\

\noindent The GHS blackhole is a member of a family of solutions
to low-energy string theory described by the action (in the string frame)
\begin{equation}
\Gamma = \int d^{4}x \sqrt{-g} e^{-2\phi}
\left[ -R -4(\nabla\phi)^2 + F^{2}\right]
\label{1}
\end{equation}
where $\phi$ is the dilaton field and $F_{\mu\nu}$ is the Maxwell field
associated with a $U(1)$ subgroup of $E_{8}\times E_{8}$ 
or ${\it{Spin}(32)}/Z_{2}$. Its charged black
hole solution is given by
\begin{equation}
ds^{2}_{string} = -f(r)dt^{2} + \frac{1}{h(r)}dr^{2} + r^{2}d\Omega
\label{2} 
\end{equation}
where,
\begin{eqnarray}
f(r) &=& \left( 1- \frac{2Me^{\phi_{0}}}{r}\right)
\left(1 - \frac{Q^{2}e^{3\phi_{0}}}{Mr}\right)^{-1}\nonumber\\
h(r) &=& \left(1-\frac{2Me^{\phi_{0}}}{r}\right)
\left(1- \frac{Q^{2}e^{3\phi_{0}}}{Mr}\right)
\label{2a}
\end{eqnarray}
with $\phi_{0}$ being the asymptotic constant value of the dilaton field.
We consider the case when $Q^{2}<2e^{-2\phi_{0}}M^{2}$
for which the above metric describes a blackhole with an event horizon 
situated at   
\begin{eqnarray}
r_{H}&=&2Me^{\phi_{0}}~.
\label{hor}
\end{eqnarray}
With the aid of dimensional reduction procedure one can effectively describe
a theory with a metric given by the by the ``$r-t$" sector of the full 
spacetime metric (\ref{2}) near the horizon. 

\noindent Now we divide the spacetime into two regions. In the region outside
the horizon the theory is free from anomaly and hence we have the 
energy-momentum tensor satisfying the conservation law
\begin{equation}
\nabla_{\mu}T^{\mu}_{(o)\nu} = 0~.\label{5}
\end{equation}
However, the omission of the ingoing modes in the 
region $r\in[r_{+}, \infty]$ near the horizon, leads to
an anomaly in the energy-momentum tensor there. As we have
mentioned earlier, in this paper we shall focus only on the
covariant form of $d=2$ gravitational anomaly given by (\cite{rw, iso}):
\begin{equation}
\nabla_{\mu}T^{\mu}_{(H)\nu} = \frac{1}{96\pi}
\bar\epsilon_{\nu\mu}\partial^{\mu}R = \mathcal A_{\nu}
\label{cov}
\end{equation}
where, $\bar\epsilon^{\mu\nu}=\epsilon^{\mu\nu}/\sqrt{-g}$
and $\bar\epsilon_{\mu\nu}=\sqrt{-g}\epsilon_{\mu\nu}$ are two
dimensional antisymmetric tensors for the upper and lower cases
with $\epsilon^{tr}=\epsilon_{rt}=1$.
It is easy to check that for the metric (\ref{2}), the 
anomaly is purely timelike with
\begin{eqnarray}
\mathcal A_{r} &=& 0\nonumber\\
\mathcal A_{t} &=& \frac{1}{\sqrt{-g}}\partial_{r}N^{r}_{t}
\label{8}
\end{eqnarray}
where,
\begin{equation}
N^{r}_{t} = \frac{1}{96\pi}\left(hf'' + 
\frac{f'h'}{2} - \frac{f'^{2}h}{f}\right).
\label{9}
\end{equation}
Now outside the horizon, the conservation equation (\ref{5})
yields the differential equation
\begin{equation}
\partial_{r}(\sqrt{-g}T^{r}_{(o)t}) = 0 \label{5a}
\end{equation}
which after integration leads to
\begin{equation}
T^{r}_{(o)t}(r) = \frac{a_{o}}{\sqrt{-g}}
\label{6}
\end{equation}
where, $a_{o}$ is an integration constant.
In the region near the horizon, the anomaly equation (\ref{cov})
leads to the following differential equation
\begin{equation}
\partial_{r}\left(\sqrt{-g}T^{r}_{(H)t}\right) = \partial_{r}N^{r}_{t}(r)
\label{900}
\end{equation}      
which after solution leads to
\begin{equation}
T^{r}_{(H)t} = \frac{1}{\sqrt{-g}}
\left(b_{H} + N^{r}_{t}(r) - N^{r}_{t}(r_{H})\right)
\label{10}
\end{equation}
where, $b_{H}$ is an integration constant. 

\noindent Now as in (\cite{iso}, \cite{rb}),  
writing the energy-momentum tensor as a sum of two contributions
\begin{equation}
{T^{r}}_{t}(r) = T^{r}_{(o)t}(r)\theta(r-r_{H}-\epsilon) 
+ T^{r}_{(H)t}(r)H(r)
\label{11}
\end{equation}
where, $ H(r) = 1 - \theta(r-r_{H}-\epsilon)$, we find 
\begin{eqnarray}
\nabla_{\mu}{T^{\mu}}_{t}
&=&\partial_{r}{T^{r}}_{t}(r)+\partial_{r}(\ln\sqrt{-g}){T^{r}}_{t}(r)
\nonumber\\
&=&\frac{1}{\sqrt{-g}}\partial_{r}(\sqrt{-g}{T^{r}}_{t}(r))\nonumber\\
&=&\frac{1}{\sqrt{-g}}\left[\left(\sqrt{-g}(T^{r}_{(o)t}(r) - T^{r}_{(H)t}(r))
+N^{r}_{t}(r)\right)\delta(r-r_{+}-\epsilon) +
\partial_{r}\left(N^{r}_{t}(r) H(r)\right)\right].
\label{12} 
\end{eqnarray}
The term in the total derivative is cancelled by quantum effects of
classically irrelevant ingoing modes. The quantum effect to cancel this
term is the Wess-Zumino term induced by the ingoing modes near the horizon. 
Hence the vanishing of the Ward identity under diffeomorphism transformation
implies that the coefficient of the delta function in the above 
equation vanishes
\begin{equation}
T^{r}_{(o)t} - T^{r}_{(H)t} + \frac{N^{r}_{t}(r)}{\sqrt{-g}} = 0~.
\label{14}
\end{equation}
Substituting (\ref{6}) and (\ref{10}) in the above equation, 
we get
\begin{equation}
a_{o} = b_{H} - N^{r}_{t}(r_{H})~.
\label{15}
\end{equation}
The integration constant $b_{H}$ can be fixed by imposing 
that the covariant energy-momentum tensor vanishes at the horizon. 
From (\ref{10}), this gives $b_{H}=0$.
Hence the total flux of the energy-momentum tensor is given by
\begin{eqnarray}
a_{o}&=&-N^{r}_{t}(r_{H})\nonumber\\
&=&\frac{1}{192\pi} f'(r_{H})g'(r_{H})~.
\label{flux}
\end{eqnarray}
Using (\ref{2a}), we finally obtain
\begin{eqnarray}
a_{o}&=&\frac{\pi}{12}T_{H}^{2}
\label{flux1}
\end{eqnarray}
		where $T_{H}$ is the Hawking temperature given by
\begin{eqnarray}
T_{H}&=&\frac{1}{8\pi Me^{\phi_{0}}}~.
\label{temp}
\end{eqnarray}
This is precisely the Hawking flux obtained in (\cite{das})
using Robinson-Wilczek method of cancellation of consistent
anomaly.   

\noindent Finally at extremality, i.e. when 
$Q^{2}=2e^{-2\phi_{0}}M^{2}$, the GHS blackhole solution
(\ref{2}, \ref{2a}) becomes
\begin{eqnarray}
ds^{2}&=&-dt^{2}+\left(1-\frac{2Me^{\phi_{0}}}{r}\right)^{-2}
+r^{2}d\Omega~.
\label{ghssol1}
\end{eqnarray}
It is easy to check that in this case the Hawking temperature
vanishes. Indeed, explicit computation of $N^{r}_{t}$ for the
above metric (\ref{ghssol1}) shows that the energy flux vanishes.\\

\noindent {\it{Hawking radiation from D1-D5 non-extremal blackhole :}}\\

\noindent As another example of covariant anomaly cancellation approach,
we consider a non-extremal five dimensional blackhole which originates
as a brane configuration in Type $IIB$ superstring
theory compactified on $S^1 \times T^4$. 
The configuration relevant to the present case is
composed of D1-branes wrapping $S^1$, D5-branes wrapping $S^1 \times
T^4$ and momentum modes along $S^1$.  The solution of the Type IIB
supergravity corresponding to this configuration is a supersymmetric
background known as the extremal five-dimensional D1-D5 blackhole
having zero Hawking temperature. Hence in order to consider Hawking radiation
we study the non-extremal  D1-D5 blackhole. 

\noindent The ten-dimensional supergravity background
corresponding to the non-extremal D1-D5 blackhole has the following
form in the string frame \cite{hms}:
\begin{eqnarray}
ds^2_{10}&=& f_1^{-1/2} f_5^{-1/2}( - h f_n^{-1} dt^2 
               + f_n ( dx_5 + (1-\tilde{f}_n^{-1}) dt)^2 )\nonumber \\
&&+ f_1^{1/2} f_5^{-1/2} (dx_6^2+ \cdots + dx_9^2)
+ f_1^{1/2} f_5^{1/2} ( h^{-1} dr^2 + r^2 d \Omega_3^2)\nonumber \\
e^{-2 \phi} &=& f_1^{-1} f_5 \quad, \quad
C_{05} = \tilde{f}_1^{-1} -1  \nonumber \\
F_{ijk} &=& \frac{1}{2} \epsilon_{ijkl} 
          \partial_l \tilde{f}_5 \quad,\quad
i,j,k,l=1,2,3,4
\label{10dsol}
\end{eqnarray}
where, $x_5$ and $x_6,\dots, x_9$ are periodic coordinates along $S^1$ and
$T^4$ respectively and $F$ is the three-form field strength 
of the RR 2-form gauge potential $C$, $F = dC$. 
Also various functions appearing in the above background
are functions of coordinates $x_1, \dots, x_4$ given by
\begin{eqnarray}
h &=& 1 - \frac{r_0^2}{r^2} \quad, \quad
f_{1,5,n} = 1+ \frac{r_{1,5,n}^2}{r^2} \nonumber \\
\tilde{f}_{1,n}^{-1} &=& 1 - \frac{r_0^2 \sinh \alpha_{1,n} 
                 \cosh \alpha_{1,n}}{r^2} f_{1,n}^{-1} \nonumber \\
r^2_{1,5,n}&=& r_0^2 \sinh^2 \alpha_{1,5,n} \quad, \quad
r^2= x_1^2 + \cdots + x_4^2
\end{eqnarray}
where, $r_0$ is the extremality parameter and $h$, $f_{1,5,n}$,
are harmonic functions representing the non-extremality and the
presence of D1, D5, and momentum modes respectively.

\noindent Dimensional reduction of (\ref{10dsol}) along $S^1 \times
T^4$ following the procedure of \cite{mss} yields the Einstein metric
of the non-extremal five-dimensional blackhole as
\begin{eqnarray}
ds^2_5 &=& -\lambda^{-2/3} h~dt^2 + 
\lambda^{1/3} (h^{-1} dr^2 + r^2 d \Omega_3^2) 
\label{5dmetric}
\end{eqnarray}
where $\lambda$ is defined by
\begin{eqnarray}			
\lambda = f_1 f_5 f_n ~.
\end{eqnarray}
The event horizon $r_H$ of this blackhole geometry is located at
\begin{eqnarray}
r_H = r_0 ~.
\label{location}
\end{eqnarray}

\noindent Apart from the metric, 
the dimensional reduction gives us three kinds
of gauge fields.  The first one is the Kaluza-Klein gauge field
$A^{(K)}_\mu$ coming from the metric and the second one, say
$A^{(1)}_\mu$, basically stems from $C_{\mu 5}$.  (We note that $\mu =
0,1,2,3,4$.) From the background (\ref{10dsol}), the first two gauge
fields are obtained as
\begin{eqnarray}
A^{(K)} &=& -(\tilde{f}_n^{-1} - 1 )dt \quad, \quad
A^{(1)} = ( \tilde{f}_1^{-1} - 1 )dt ~.
\label{gauge}
\end{eqnarray}
Unlike these gauge fields which are one-form in nature, the last one
is the two-form gauge field $A_{\mu\nu}$, originating from
$C_{\mu\nu}$ whose field strength is given by the expression of $F$
in (\ref{10dsol}).

\noindent Now if we consider a free complex scalar field in the black
hole background (\ref{5dmetric}) and (\ref{gauge}) and
perform a partial wave decomposition of $\varphi$
in terms of the spherical harmonics, then it can be shown
that the action near the horizon becomes \cite{D1}
\begin{eqnarray}
S[\varphi]&=&-\sum_a \int dtdr~ r^3~\lambda^{1/2}~
  \varphi^\star_{a}(t, r) 
  \left( - \frac{1}{f} (\partial_t -i A_t)^2 
          + \partial_r f \partial_r 
  \right)
  \varphi_{a}(t, r) 
\label{action}
\end{eqnarray}
where, $A_t = e_1 A^{(1)}_t + e_K A^{(K)}_t$ and 
$a$ is the collection of angular quantum numbers
of the spherical harmonics. It can be easily checked that
this action describes an infinite set of massless
two-dimensional complex scalar fields in the following background :
\begin{equation}
ds^2 = -f(r)dt^2 + \frac{1}{f(r)}dr^2\quad,
\quad \Phi = r^3 \lambda^{1/2}
\label{2.1}
\end{equation} 
\begin{equation}
A_{t}(r) = - \frac{e_{1}r_{o}^2 \sinh \alpha_{1}
\cosh\alpha_{1}}{r^2 + r_{o}^2}
+ \frac{e_{k}r_{o}^2 \sinh \alpha_{n}
\cosh\alpha_{n}}{r^2 + r_{n}^2}
\label{2.2}
\end{equation}
where $\Phi$ is the two-dimensional dilaton field.

\noindent As we have stated earlier, 
the two dimensional effective theory near
the horizon (\ref{action}) possesses gravitational 
as well as gauge anomaly. We once again follow the approach
based on covariant anomaly cancellation proposed in \cite{rb}. 
We first consider the gauge
part. Since there are two kinds of $U(1)$ gauge symmetries, we have 
two $U(1)$ gauge currents $J_{\mu}^{(1)}$ and $J_{\mu}^{(K)}$ corresponding
to $A_{\mu}^{(1)}$ and $A_{\mu}^{(K)}$ respectively. The form for
covariant gauge anomaly for these two currents are identical
in nature, therefore we discuss the case for $J_{\mu}^{(1)}$
explicitly and just mention the result for the other. Since the spacetime
has been divided into two regions, we divide the 
current $J^{(1)\mu}$ into two parts. The current 
outside the horizon denoted by $J^{(1)\mu}_{(o)}$ 
is anomaly free and hence satisfies the conservation law
\begin{equation}
\nabla_{\mu}J^{(1)\mu}_{(o)} = 0 \label{2.4}  
\end{equation} 
while the current near the horizon satisfies 
\begin{eqnarray}
\nabla_{\mu}J^{(1)\mu}_{(H)}&=& -\frac{e_1}{4\pi}
\bar\epsilon^{\rho\sigma}F_{\rho\sigma}= 
\frac{e_1}{2\pi}\partial_{r}A_{t}~.\label{2.3}  
\end{eqnarray}  
Solving (\ref{2.4}) and (\ref{2.3}) in the region outside and near
the horizon, we get
\begin{eqnarray}
J^{(1)r}_{(o)} &=& c_{o}^{(1)}\label{2.5}\\
J^{(1)r}_{(H)} &=& c_{H}^{(1)} +\frac{e_1}{2\pi}
\left[A_{t}(r) - A_{t}(r_{H})\right].\label{2.6}
\end{eqnarray}
Now as in \cite{iso}, writing $J^{(1)r}$ as 
\begin{equation}
J^{(1)r} = J^{(1)r}_{(o)}\Theta(r-r_{H} -\epsilon) + J^{(1)r}_{(H)} H(r)
\label{2.7}
\end{equation}
we find 
\begin{equation}
\nabla_{\mu}J^{(1)\mu} = \partial_{r}J^{(1)r} = 
\partial_{r}\left(\frac{e_{1}}{2\pi}A_{t}H\right) 
+ \delta(r-r_{H}-\epsilon)\left[J^{(1)r}_{(o)} 
- J^{(1)r}_{(H)} +\frac{e_1}{2\pi}A_{t} \right].
\label{2.8}
\end{equation}
Now the vanishing of the Ward identity under gauge transformation 
requires that the first term must be cancelled 
by quantum effects of classically irrelavent ingoing modes. 
The vanishing of the second term implies that the 
coefficient of the delta function is zero, leading
to the condition
\begin{equation}
c^{(1)}_{o} = c^{(1)}_{H} - \frac{e_1}{2\pi}A_{t}(r_{H})~. 
\label{2.9}
\end{equation}                 
The coefficient $c^{(1)}_{H}$ vanishes by requiring that 
the covariant current $J^{(1)r}_{(H)}$ 
vanishes at the horizon. Hence the charge flux
corresponding to $J^{(1)r}$ is given by  
\begin{equation}
c^{(1)}_{o} = -\frac{e_1}{2\pi}A_{t}(r_{H}) = 
\frac{e_1}{2\pi}(e_1\tanh \alpha_1 - e_K \tanh \alpha_n)~.
\label{2.10}
\end{equation}
Following the same procedure for $J^{(K)\mu}$ 
satisfying the anomaly equation
\begin{eqnarray}
\nabla_{\mu}J^{(K)\mu}_{(H)}&=& -\frac{e_K}{4\pi}
\bar\epsilon^{\rho\sigma}F_{\rho\sigma}= 
\frac{e_K}{2\pi}\partial_{r}A_{t}\label{2.30aa}  
\end{eqnarray}  
we find that the charge flux
corresponding to current $J^{(K)r}$ reads
\begin{equation}
c^{(K)}_{o} = -\frac{e_K}{2\pi}A_{t}(r_{H}) = 
\frac{e_K}{2\pi}(e_1 \tanh \alpha_1 - e_K \tanh \alpha_n)~.
\label{2.11}
\end{equation}
Hence the total charge flux is given by
\begin{equation}
c_{o} = c^{(1)}+c^{(K)}_{o} = -\frac{e}{2\pi}A_{t}(r_{H}) 
= \frac{e}{2\pi}(e_1 \tanh \alpha_1 - e_K \tanh \alpha_n)~.
\label{2.12}
\end{equation}
where, $e = e_1 + e_K$.

%%%%%%%%%%%%%%%%%%%%%%%%%%%%%%%%%%%%%%%%%%%%%%%%%%%%%%%%%%%%%%
\noindent Now we move on to the problem of computing the energy flux.
Since we have an external gauge field,
the energy-momentum tensor will not satisfy the conservation law 
even at classical level, rather
it gives rise to the Lorentz force law, 
$\nabla_{\mu}{T^{\mu}}_{\nu} = F_{\mu\nu}J^{\mu}$
\footnote{Note that here $J^{\mu}=J^{(1)\mu}+J^{(K)\mu}$.}. 
Hence the corresponding expression for the anomalous 
Ward identity for covariantly
regularised quantities is given by \cite{rb} 
\begin{equation}
\nabla_{\mu}T^{\mu}_{\quad\nu} = F_{\mu\nu}J^{\mu} + \mathcal{A}_{\nu}
\label{2.13}
\end{equation} 
where, $\mathcal{A}_{\nu}$ is the two-dimensional 
gravitational covariant anomaly (\ref{cov}).
In the region outside the horizon, 
there is no anomaly and hence the Ward identity reads
\begin{equation}
\nabla_{\mu}{T^{\mu}}_{(o)\nu} = \partial_{r}{T^{r}}_{(o)t} 
= F_{rt}J^{r}_{(o)}
\label{2.14}
\end{equation}
Using (\ref{2.5}), the above equation can be solved as
\begin{eqnarray}
T^{r}_{(o)t}(r)&=& a_{o} + c_{o}A_{t}(r)
\label{z3}
\end{eqnarray}
where, $a_{o}$ is an integration constant. However near the horizon 
the Ward identity reads
\begin{equation}
\partial_{r}T^{r}_{(H)t} = F_{rt}J^{r}_{(H)} + \partial_{r}N^{r}_{t}
\end{equation}
where, $N^{r}_{t}$ is given by (\ref{9}) with $h(r)=f(r)$.
Now substituting $J^{r}_{(H)}=J^{(1)r}_{(H)}+J^{(K)r}_{(H)}$, we get
\begin{equation}
T^{r}_{t(H)} = a_{H} + \int^{r}_{r_{H}} dr 
~\partial_{r}\left[c_o A_{t}+
\frac{e}{4\pi}A_{t}^2 +  N^{r}_{t}\right]. 
\end{equation}
Now following the same procedure 
as given in the gauge part we arrive at
the relation
\begin{equation}
a_o = a_H + \frac{e}{4\pi}A_{t}^2(r_{H}) - N^{r}_{t}(r_{H})~. 
\end{equation} 
Implimenting the boundary condition that covariant energy momentum 
tensor vanishes at the horizon as before fixes $a_{H}$ to be zero.
Therefore, $a_{o}$ is given by
\begin{equation}
a_{(o)} =  \frac{e}{4\pi}A_{t}^2 (r_{H}) - N^{r}_{t}(r_{H})~.
\end{equation}
Now substituting the values for $A_{t}$ and 
$N^{r}_{t}$ at the horizon, we get
\begin{eqnarray}
a_o &=& \frac{e}{4\pi}A_t^2(r_H) +N^r_t(r_+) \nonumber\\
&=& \frac{e}{4\pi} (e_1 \tanh \alpha_1 - e_K \tanh \alpha_n)^2
 +\frac{\pi}{12} T_H^2 
\label{tflux}
\end{eqnarray}
where,
\begin{eqnarray}
T_H&=&\frac{1}{2\pi r_{0}\cosh\alpha_{1}\cosh\alpha_{5}\cosh\alpha_{n}}~.
\label{temp11}
\end{eqnarray}
This is just the energy flux from blackbody radiation with two chemical
potentials for the charges $e_{1}$ and $e_{K}$ \cite{D1}.
  
%\begin{eqnarray}
%T^{r}_{(H)t} &=& b_{H} + N^{r}_{t}(r) - N^{r}_{t}(r_{H})
%\label{z1}
%\end{eqnarray}
%\begin{eqnarray}
%T^{r}_{(o)t} - T^{r}_{(H)t} + N^{r}_{t}(r)&=& 0
%\label{z2}
%\end{eqnarray}
%from which we again get
%\begin{equation}
%a_{o} = b_{H} - N^{r}_{t}(r_{H})~.
%\label{z4}
%\end{equation}

%%%
%\vskip 0.5cm
%\noindent Note added : We would like to mention that on the
%day we submitted our paper to the arXiv, a similar paper appeared
%there (S. Q. Wu et.al, e-print: arXiv 0707.2449, [hep-th]). 
%However, as an example of our procedure we
%consider the GHS blackhole 
%which is not discussed in their paper.
%%%%%%%%%%%%%%%%%%%%%%%%%%%%%%%%%%%%%%%%%%%%%%%%%%%%%%%%%%%%%%%%%%
%%%%%%%%%%%%%%%%%%%%%%%% Discussions %%%%%%%%%%%%%%%%%%%%%%%%%%%%%%%
\noindent {\it{Discussions :}}

\noindent In this paper, we studied the problem of Hawking radiation 
from blackhole spacetimes that occur in 
string theory using covariant anomaly cancellation technique
proposed in \cite{rb}. 
The point is that Hawking radiation plays
the role of cancelling gauge and gravitational
anomalies at the horizon to restore the gauge/diffeomorphism
symmetry at the horizon.
An advantage of this method is that neither
the consistent anomaly nor the counterterm relating the different
(covariant and consistent) currents, 
which were essential ingredients in \cite{iso, das, D1},
were required. 

\noindent We discussed in particular Hawking radiation
from GHS and five dimensional non-extremal D1-D5 blackhole.
For GHS blackhole, the energy-momentum flux 
was obtained when $Q^2 < 2Me^{-2\phi_{o}}$.
At extremality there is no energy flux and hence Hawking temperature
is zero. In the case of $D1-D5$ blackhole, fluxes of electric charge flow 
and energy-momentum tensor were obtained. The resulting fluxes 
are the same as that of the two dimensional black body 
radiation at the Hawking temperature.

%%%%%%%%%%%%%%%%%%%%%%%%%%%%%%%%%%%%%%%%%%%%%%%%%%%%%%%%%%%%%%%%%%
\section*{Acknowledgements }
The authors would like
to thank Prof. R. Banerjee for useful comments and constant
encouragement.

%%%%%%%%%%%%%%%%%%%%%%%%%%%%%%%%%%%%%%%%%%%%%%%%%%%%%%%%%%%%%%%%%%

\end{document}